\documentclass[Article]{latex/emulateapj}

\usepackage{epsfig}
\usepackage{xspace}
\usepackage{amsmath, amsfonts, amssymb, cancel, graphicx, hyperref, color, xcolor, float, esint}

\slugcomment{Submitted to JCAP}

\shorttitle{Constraining halo thermodynamics with future SZ observations}
\shortauthors{Battaglia et al.}

\newcommand{\be}{\begin{equation}}
\newcommand{\ee}{\end{equation}}
\newcommand{\bea}{\begin{eqnarray}}
\newcommand{\eea}{\end{eqnarray}}
\newcommand{\bal}{\begin{aligned} }
\newcommand{\eal}{\end{aligned}}
\newcommand{\rmn}{\mathrm}
\newcommand{\te}{\tau}
\newcommand{\st}{\sigma_\mathrm{T}}
\newcommand{\me}{m_\mathrm{e}}
\newcommand{\nume}{n_\mathrm{e}}
\newcommand{\dd}{\mathrm{d}}

\newcommand{\eps}{\epsilon}
\newcommand{\nth}{\alpha}

\begin{document}

\title{Future constraints on halo thermodynamics from combined Sunyaev-Zel'dovich measurements}

\author{Nicholas~Battaglia$^{1}$, Simone~Ferraro$^{2,3}$, Emmanuel~Schaan$^{1}$, David~N.~Spergel$^{1,4}$}

\altaffiltext{1}{Dept. of Astrophysical Sciences, Princeton University, Princeton, NJ 08544, USA}
\altaffiltext{2}{Berkeley Center for Cosmological Physics, and Dept. of Astronomy, University of California, Berkeley, CA  94720, USA}
\altaffiltext{3}{Miller Institute for Basic Research in Science, University of California, Berkeley, CA, 94720 USA}
\altaffiltext{4}{Center for Computational Astrophysics, Flatiron Institute, 162 Fifth Avenue, New York, NY 10010, USA}

\begin{abstract}

The improving sensitivity of measurements of the kinetic Sunyaev-Zel'dovich (SZ) effect opens a new window into the thermodynamic properties of the baryons in halos. We propose a methodology to constrain these thermodynamic properties by combining the kinetic SZ, which is an unbiased probe of the free electron density, and the thermal SZ, which probes their thermal pressure. We forecast that our method constrains the average thermodynamic processes that govern the energetics of galaxy evolution like energetic feedback across all redshift ranges where viable halos sample are available. Current Stage-3 cosmic microwave background (CMB) experiments like AdvACT and SPT-3G can measure the kSZ and tSZ to greater than 100$\sigma$ if combined with a DESI-like spectroscopic survey. Such measurements translate into percent-level constraints on the baryonic density and pressure profiles and on the feedback and non-thermal pressure support parameters for a given ICM model. This in turn will provide critical thermodynamic tests for sub-grid models of feedback in cosmological simulations of galaxy formation. The high fidelity measurements promised by the next generation CMB experiment, CMB-S4, allow one to further sub-divide these constraints beyond redshift into other classifications, like stellar mass or galaxy type.
\end{abstract}

  
\section{Introduction}

Star formation in galaxies is inefficient: less than 10 \% of the available baryons today are turned into stars \citep[e.g.,][and references therein]{Gallazzi2008}. This raises a fundamental question for theoretical models of galaxy formation and evolution: what physical processes cause global star formation to be so inefficient? Fiducial theoretical models for galaxy formation, both semi-analytic \citep[e.g.,][]{Kauf1993,Croton2006,RS2015} and simulations based \citep[e.g.,][]{illustris,Eagles,HorizonAGN,Fire2}, invoke feedback processes from massive stars, supernovae and active galactic nuclei (AGN). These inject additional energy into the intrastellar, intergalactic, or intracluster mediums (ISM, IGM and ICM, respectively). In order to understand galaxy evolution and star formation, the physical processes and thermodynamic properties that govern the baryons in galaxies and clusters need to be understood and measured. These open questions on the formation and evolution of galaxies and galaxy clusters are important for future cosmological probes (e.g., late-time growth of structure measurements) which push into the quasi-linear and non-linear regime where modeling such baryonic effects cannot be ignored. \citep[e.g.,][]{vD2011,Semboloni2011,Eifler2015}.

The improvements in the sensitivity of high-resolution CMB experiments, such as the Advanced Atacama Cosmology Telescope \citep[AdvACT;][]{AdvACT} the South Pole Telescope-3G \citep[SPT-3G;][]{SPT3G} are opening up a new window on baryonic physics through measurements of the Sunyaev Zel'dovich effect \citep[SZ,][]{SZ1970,SZ1972}. As they propagate through the Universe, a small fraction of the CMB photons are Thomson-scattered by free electrons in the intergalactic medium (IGM) and intracluster medium (ICM).
Two important resulting effects are the thermal and kinetic SZ effects. The thermal SZ is the increase in energy of CMB photons due to scattering with hot electrons, which imprints a unique spectral distortion in the CMB blackbody that is a decrement in thermodynamic temperature at frequencies below 217 GHz and excess at higher frequencies. The kinetic SZ is the result of CMB photons Thomson-scattering off free electrons that have a non-zero peculiar velocity with respect to the CMB rest frame, which produces small shifts in the CMB temperature. Both the tSZ and kSZ effects are direct probes of the thermodynamic properties of the IGM and ICM since their magnitudes are proportional to the integrated electron pressure (tSZ) and momentum (kSZ) along the line-of-sight. Therefore, the tSZ and kSZ probe the IGM and ICM radial pressure and density profiles, which are sensitive to baryonic processes like star-formation, feedback, and non-thermal pressure support. These measurements, combined with the gravitational potential profile measurements from weak lensing, are sufficient to solve the thermodynamic properties of the halo. 

For example, we will derive a virial equation that can measure both the thermal and non-thermal pressure support, $P_\rmn{Nth}$:

\be
\underbrace{ \frac{1}{3}\Phi_\text{gas} } 
_{\substack{\text{from kSZ} \\ \& \text{mass profile}}}
 + \underbrace{\int P_\rmn{th} \dd V }_\text{from tSZ} + \underbrace{\int P_\rmn{Nth} \dd V}_{\rightarrow \text{inferred}} \propto P_\rmn{Surface}, \nonumber
\ee

\noindent here $\phi_\rmn{gas}$ is analogous to the gravitational energy of the gas, $P_\rmn{th}$ is the thermal pressure, and $P_\rmn{Surface}$ is a surface pressure (See the Appendix for a full derivation).

Since the initial detections of the tSZ effect on individual galaxy clusters \citep[e.g.,][]{Parl1972,Meyer1983,Part1987}, there now exists an abundance of tSZ measurements across a large range of halo masses, from galaxy clusters mass-scales where we have catalogs of tSZ detected clusters \citep[e.g.,][]{Reic2013,Hass2013,Bleem2015,PlanckCat2016}, down to galaxy mass-scales where the tSZ is detected via stacked observations \citep[e.g.,][]{Hand2011,Planckstack,Greco2015,Spacek2016,Spacek2017}. The kSZ effect was first detected \citep{Hand2012} recently using a galaxy catalog from the Baryon Oscillation Spectroscopic Survey \citep{BOSS2011} and CMB observations from the ACT telescope. Since then, many other detections from multiple CMB experiments using various techniques and galaxy catalogs have followed \citep{PlanckkSZ,Schaan2016,Hill2016,SPTkSZ2016,FDB2016}, as well as detections from individual galaxy clusters \citep{Sayers2013,Adam2017}. While all the current detections are around 4-$\sigma$, forecasts for experiments like AdvACT, SPT-3G, the Simons Observatory, and CMB Stage-4 \citep[CMB-S4,][]{CMBS4} promise an order magnitude and greater improvements from these current detections \citep{Flender2016,F16}.

In this paper, we propose to use stacked observations of tSZ and kSZ to measure the average density and pressure profiles for a given galaxy, quasar, or cluster sample. We show that these measurements place constraints on the important physical processes in the IGM and ICM, such as AGN feedback and non-thermal pressure support, which impact the baryons that govern global star formation and thus, galaxy evolution. Previous work proposed to use joint tSZ and kSZ measurements to infer the optical depth, temperature and peculiar velocity of individual clusters \citep{Knox2004,Sehgal2005}, whereas in this work we focus on baryonic processes for ensemble populations.

We provide theoretical formalism for interpreting combined SZ observations and inferring parametric and non-parametric constraints on galaxy formation in Section \ref{sec:meth}. In Section \ref{sec:stats} we describe the component separation and observational techniques used in this work. We present the forecasts for our proposed method in Section \ref{sec:res}. We discuss possible extension of this work and conclude our findings in Section \ref{sec:con}. We adopt a flat $\Lambda$CDM cosmology with $\Omega_{\rm M}=0.25$, $H_0=70$ km~s$^{-1}$~Mpc$^{-1}$. The results presented in this work are not sensitive to cosmological parameters.

\section{Methodology}
\label{sec:meth}

The tSZ spectral distortion and kSZ doppler boost amplitudes are proportional to the line-of-sight pressure and momentum, respectively. For the tSZ, the distortion is a function of frequency and the Compton $y$ parameter: 

\be
\frac{\Delta T(\nu)}{T_\rmn{CMB}} = f(\nu) y, 
\ee

\noindent where $T_\rmn{CMB}$ is the CMB temperature, $f(\nu) = x\,\rmn{coth}(x/2) - 4$, $x = h\nu / (k_B T_\rmn{CMB})$, $h$ is the Planck constant, and $k_B$ is the Boltzmann constant. Here we neglected the relativistic corrections to $f(\nu)$ \citep[e.g.,][]{Nozawaetal2006,Chluba2012}. The Compton-$y$ parameter is defined as

\be
y = \frac{\st}{\me c^2} \int_\rmn{LOS} P_\rmn{e} \ \dd l,
\label{eq:y}
\ee

\noindent where $\st$ is the Thomson cross-section, $c$ is the speed of light, $\me$ is the electron mass, $P_\rmn{e}$ is the thermal electron pressure and $\dd l$ is the line-of-sight (LOS) physical distance.

The kSZ effect is sensitive to the combination of optical depth and peculiar velocity, $v_r$ of each halo along the line-of-sight, 

\be
\frac{\Delta T}{T_\rmn{CMB}} = \frac{\st}{c} \int_\rmn{LOS} e^{-\tau} \nume v_r \ \dd l, 
\ee

\noindent  where $\nume$ is the electron number density and the optical depth $\tau$ is defined as

\be
\te = \st \int_\rmn{LOS} \nume \dd l.
\label{eq:tau}
\ee

The ICM is observed to have entropically stratified medium and its temperature and pressure are mostly determined by its gravitational potential \citep[e.g.,][]{Cav2009,Walker2012,Eckert2013}. We assume these properties hold for the IGM in groups and massive galaxies, which is observed in local massive galaxies \citep[e.g.,][]{Forman1985,Hump2006,DS2007,Voit2015,AG2016} and is seen in simulations of massive galaxies \citep[e.g.,][]{Freeke2016} and groups \citep[e.g.,][]{LeBrun2014}. In steady state, the total pressure of the hot gas (whether thermal and non-thermal) balances the gravitational force on it. For a spherically symmetric system, this reads as:

\be
\frac{\dd P_\rmn{tot}}{\dd r} = - G\frac{M(<r)}{r^2} \rho_\rmn{gas}(r).
\label{eq:HE}
\ee
Here $G$ is the gravitational constant, $M(<r)$ is the total mass enclosed within radius $r$, $\rho_\rmn{gas}$ is the ICM gas density, $P_\rmn{tot}$ is the total ICM pressure in the system which can be operated into thermal ($P_\rmn{th}$) and non-thermal ($P_\rmn{Nth}$) contributions such that $P_\rmn{tot} = P_\rmn{th} + P_\rmn{Nth}$. Observations of tSZ and kSZ measure $P_\rmn{th}$ and $\rho_\rmn{gas}$ directly and we can measure $M(<r)$ using optical weak lensing \citep[e.g.,][]{Rachel2006,Alexie2016} or use an NFW model \citep{NFW1997} for the dark matter (DM) component. The remaining $P_\rmn{Nth}$ profile can be constrained non-parametrically by these cross-correlations or both feedback and non-thermal pressure support can be constrained if one invokes a parametric model for the ICM.

The projected measurements for tSZ and kSZ are related to the three-dimensional radial profiles $P_\rmn{e}$ and $\nume$ by
\be
y (\theta_d) = \frac{\st}{\me c^2}  \int_\rmn{LOS} P_\rmn{e} \left(\sqrt{l^2 + d^2_A(z)|\theta_d |^2 } \right) \dd l,
\label{eq:yproj}
\ee

\noindent and 

\be
\tau (\theta_d) = \st \int_\rmn{LOS} \nume \left(\sqrt{l^2 + d^2_A(z)|\theta_d |^2 } \right) \dd l,
\label{eq:tauproj}
\ee

\noindent where $r^2 = l^2+d^2_A(z)|\theta|^2$ and $d_A(z)$ is the angular diameter distance to redshift $z$. Equations \ref{eq:yproj} and \ref{eq:tauproj} are used to model the observations and are convolved with the beams for each experimental and observational setup. We convert $P_\rmn{e} = (2 + 2X_\rmn{H}) / (3 + 5X_\rmn{H}) P_\rmn{th}$ assuming a fully ionized medium of primordial abundances, where $X_\rmn{H}$ is the hydrogen mass fraction ($X_\rmn{H} = 0.76$). The same conversion is applied to $\nume$.

\subsection{Parametric IGM and ICM Model}
 
In this section, we implement a model of the IGM and ICM proposed by \citet{OBB2005}, \citet{BOV2009}, and \citet{Shaw2010}, in order to constrain the non-thermal pressure and the energy injected in the gas by feedback. This model assumes that the ICM has the same initial density and temperature structure as the DM halo, then is rearranged into pressure equilibrium with the total gravitational potential, assuming a polytropic equation of state. The DM halo is assumed to have an NFW density profile \citep{NFW1997}

\be
\rho_\rmn{DM}(x) = \frac{\rho_0}{x(1 + x)^2},
\ee

\noindent where $x \equiv r / r_s$ and $r_s$ is the scale radius of the NFW profile and $\rho_0$ is the density normalization.
The definition of the scale radius is $r_{s} \equiv R_{200}/c_{\rmn{NFW}}$ where $c_{\rm{NFW}}$ is the concentration and $R_{200}$ is a halo centric radius enclosing a mass $M_{200}$ equal to 200 times the critical density at the halo redshift, $\rho_\rmn{cr}(z) \equiv 3 H_0^2 \left( \Omega_\rmn{M}(1+z)^3 + \Omega_{\Lambda} \right) / (8\pi\,G)$.
We use a fitting formula for $c_{\rm{NFW}}$ from \citet{Duffy2008}. In principle $c_{\rm{NFW}}$ can be marginalized over or fit from a weak lensing measurement.

We allow some fraction of the initial ICM gas to cool and form stars.
The amount of stars that form is modeled as a function of the halo mass following the observed scaling relation from \citet{Gio2009},

\be
f_\star = 2.5\times10^{-2} \left(\frac{M_\rmn{200}}{7\times10^{13} M_\odot}\right)^{-0.37}.
\ee

\noindent The normalization of the scaling reaction here differs from \citet{Gio2009}, due to the difference in mass definitions. In practice, stellar mass estimates are available for the halo catalogs we propose to use, thus avoiding the uncertainty in this observed stellar-to-halo mass scaling. This is especially important when applying this model to higher redshift samples since the scaling relation presented here is calibrated close to $z=0$.  Here we choose this scaling relation \citet{Gio2009} for simplicity rather than marginalize over the parameters \citep[see][]{Flender2016}.

The final ICM density $\rho (r)$ and total pressure $P_\rmn{tot} (r)$ profiles are assumed to follow a polytropic equation of state such that,

\be
\rho_\rmn{gas} (r)= \rho_0\theta(r)^{\frac{1}{\Gamma -1}},
\ee

\noindent and

\be
P_\rmn{tot} (r)= P_0\theta(r)^{\frac{1}{\Gamma -1} + 1}.
\label{eq:ptot}
\ee

\noindent Here $\Gamma$ is the polytropic index, and $P_0$ and $\rho_0$ are the central pressure and density, respectively. The function $\theta(r)$ is referred to as the polytropic variable, defined as

\be
\theta(r) = 1 + \frac{\Gamma -1}{\Gamma} \frac{\rho_0}{P_0}(\Phi_0 - \Phi(r)),
\ee

\noindent where $\Phi_0$ is the central potential of the halo. The non-thermal pressure contribution is modeled following the fitting function proposed in \citet{Shaw2010} and validated by simulation results \citep[e.g.,][]{BBPS1,Nelson2014}. The form is 

\be
P_\rmn{Nth} = \nth  (r/R_{200})^{n_\rmn{Nth}},
\label{eq:pnth}
\ee

\noindent where the power-law radial dependence $n_\rmn{Nth}$ is fixed at 0.8 as proposed in \citet{Shaw2010} and the normalization of the non-thermal pressure support, $\nth$, is a parameter that we fit for. In what follows, we fit for $\nth$ as a function of redshift, rather than assuming a fixed redshift-dependence. Finally, we calculate the thermal pressure using $P_\rmn{th} = P_\rmn{tot} - P_\rmn{Nth}$, where $P_\rmn{tot}$ and $P_\rmn{Nth}$ are calculated using Equations~\ref{eq:ptot} and~\ref{eq:pnth}, respectively.

We solve for $P_0$ and $\rho_0$ by imposing conservation of energy, in which the final energy $E_\rmn{f}$ is related to the initial energy $E_\rmn{i}$ by 

\be
E_\rmn{f} = E_\rmn{i} + \eps M_\star c^2 + \Delta E_\rmn{p}.
\ee

\noindent Here the term $\epsilon M_\star c^2$ accounts for feedback for star formation including any feedback from central active galactic nuclei, where $\eps$ is a dimensionless efficiency parameter that couples star formation to the feedback energy, and $\Delta E_\rmn{p}$ is the work received from the surface pressure when the boundary of halo changes. We apply the boundary condition that the final total pressure at the radius of the halo is equal to the initial surface pressure of the gas
\be
P_\rmn{s,i} = P_\rmn{tot}(R_f).
\ee
Here the initial energy, $E_\rmn{i}$, and surface pressure, $P_\rmn{s,i}$, are calculated from the initial density and temperature structure according to the NFW DM profile. Additionally, we conserve the total gas mass in the halo, taking into account the mass that was converted into stars. The details of these boundary conditions can be found in the original modeling papers \citep{OBB2005,BOV2009,Shaw2010}.

The assumption of spherical symmetry in this model is sufficient for both the average gas properties \citep[e.g.,][]{BBPS1} and DM \citep[e.g.,][]{Corless2007,Becker2011}, since we are modeling stacked profiles. Additionally, hydrodynamic simulations of individual halos (massive galaxy and cluster mass scales) were shown to have projected images for the density and pressure of the inoized gas with low ellipticity \citep[e.g.,][]{Lau2011,BBPS2,Freeke2016}. We justify the assumption that the polytropic index $\Gamma$ is a constant across the halo radius with cosmological hydrodynamic simulations that show $\Gamma$ varies by less than 10\% within the virial radius \citep{BBPS2}. The resulting profiles from a similar model for the ICM have been compared to and were in agreement with average profiles from full cosmological hydrodynamic simulations \citep{Bode2012}.

Recently \citet{Flender2016b} proposed to use X-ray observations to constrain the parameters of similarly modified \citet{OBB2005} ICM model. The focus of their work was on inferring the optical depth of clusters to CMB photons and they discuss uncertainties associated with such an analysis. The constrains on the model ICM parameters in \citet{Flender2016b} are a complementary approach to this work, since their method uses X-ray observations. Furthermore \citet{KG2016} recently demonstrated the ability of thermal SZ observations to constrain non-thermal pressure support in the Coma cluster.

We stress that the model presented above is idealized, and it was chosen such that we could calculate how the density and pressure profiles depend on models of feedback and non-thermal pressure support. The ability to constrain feedback and non-thermal pressure in this method comes from the observed profiles and not the actual model. Here this model is meant to illustrate what information is accessible through this method and future measurements. We stress that when applying this method to observational data, more complicated or even full hydrodynamic simulations of large-scale structure can and should be used to infer information of feedback and non-thermal pressure support. Increasing the complexity of the model will not significantly impact the constraints shown in this work, unless more free parameters are introduced. As we will show in Section~\ref{sec:res} future experiments have the statistical power to extract information on feedback and non-thermal pressure support, which further motivates the need and importance of realistic cosmological simulations.

\subsection{Non-parametric constraints}
\label{sec:npconst}
Joint measurements of thermal pressure (tSZ), density (kSZ) and gravitational potential (lensing) allow to solve for the amount of non-thermal pressure in a model-independent way. The assumption of balance between gravity and total pressure support (whether thermal or non-thermal) can be translated into a virial theorem. Indeed, multiplying both sides of Equation~\ref{eq:HE} by $r$ and integrating over volume yields:
\be
\Phi_\text{gas} + 3 \int_0^{R} P_\text{tot}(\vec{r}) \dd^3 \vec{r} =  4\pi R^3 P_\text{tot} (R),
\label{eq:vir}
\ee
where $\Phi_\text{gas}$ is analogous to the gravitational energy of the gas in the total gravitational potential:
\be
\Phi_\text{gas} \equiv -\int_0^{R}  G\frac{M(<r)}{r}\rho_\rmn{gas} \dd^3 \vec{r} .
\ee
Note that $\Phi_\text{gas}$ would be rigorously the gravitational energy of the gas only if the gas were self-gravitating, which is not quite the case since the dark matter makes an important contribution to the mass (see Appendix~\ref{app:virial} for more details). Regardless, this distinction is of little importance for our purpose, as we explain shortly.

Further splitting the thermal and non-thermal components of the total pressure finally gives:
\be
\underbrace{ \frac{1}{3}\Phi_\text{gas} } 
_{\substack{\text{from kSZ} \\ \& \text{mass profile}}}
+ \underbrace{\int_0^{R} P_\text{th} \dd^3 \vec{r}}
_\text{from tSZ}
+ \underbrace{\int_0^{R} P_\text{Nth} \dd^3 \vec{r} }
 _{\rightarrow\text{inferred}}
 =  
\underbrace{ \frac{4}{3}\pi R^3 P_\text{tot} (R) }
_\text{external pressure}.
\label{eq:virial_full}
\ee
This Virial theorem shows the balance between thermal and non-thermal support on the one hand, and gravity and the external pressurization on the other hand. 

The quantity $\Phi_\text{gas}$ can be inferred from the gas density profile (measured from kSZ) and the total mass distribution (assumed to follow a NFW profile or measured from lensing). The volume integral of the thermal pressure support is directly inferred from the tSZ measurement, modulo the conversion from $P_\rmn{th}$ to $P_\rmn{e}$. The external pressure term can be modeled in terms of the mass accretion onto the cluster \citep[for example see][]{OBB2005}.

As a result, Equation~\ref{eq:virial_full} allows to infer the volume-averaged non-thermal pressure support in the cluster, from kSZ and tSZ measurements. This determination of the non-thermal pressure support is model-independent, and is valuable for several reasons. First, knowing the ratio of thermal to non-thermal pressure allows to constrain the hydrostatic mass bias, which affects the cluster mass-observable relation and thus constitutes a limiting systematic in cluster cosmology. Second, it may allow to quantify the amount of energy injected through feedback, by comparing the tSZ and kSZ signals from halos with different properties (e.g. absence or presence of a quasar). From tSZ measurements only, one can quantify the amount of thermal energy injected; our method allows to quantify both the thermal and non-thermal energy injection.

\section{Statistical Tools}
\label{sec:stats}
\subsection{SZ estimators}

The tSZ estimator that we forecasted for is a standard filtering process on a component separated $y$-map, the details of which are described in the subsections below. For the kSZ many estimators exist in the literature, here we forecasted on the estimator which uses reconstructed velocity fields from spectroscopic sample of galaxies to cross correlate with CMB maps \citep{Ho2009,Li2014}. Because a halo is equally likely to be moving towards us than away from us, stacking the CMB temperature map at the positions of halos results in a cancellation of the kSZ signal. The estimator we use circumvents this issue by weighting each halo by an estimate of its LOS velocity, reconstructed from the density field.

The kSZ has been detected at $> 3\sigma$ significance using this velocity reconstruction estimator on Planck and ACT observations with different spectroscopic samples \citep{PlanckkSZ,Schaan2016} and near future detection will improve on this significance greatly. The impact of small-scale foregrounds is one of the largest uncertainties in these forecasts and will be addressed in the next section.

\subsection{Component separation}
\label{sec:comp}
In order to model the impact of foregrounds on kSZ detection and the noise on the reconstructed $y$-map, we employ a simple Internal Linear Combination (ILC) technique (\cite{2004ApJ...612..633E, 2014A&A...571A..21P, 2014JCAP...02..030H, 2011MNRAS.410.2481R}).
We assume that the temperature observations $T_i$ at a frequency labeled by the index ``$i$'' in a pixel $p$ is given by a linear combination of a CMB map (which includes kSZ) $s(p)$, a tSZ map $y(p)$ and noise $n(p)$, which includes both detector noise and all of the other foregrounds:
\be
T_i(p) =  a_i s(p) + b_i y(p) + n_i(p)
\label{eq:ILCdef}
\ee
The vector $\vec{a}$ quantifies the frequency dependence of the CMB, which is just constant in CMB temperature units that we use throughout the paper. Therefore we can just take $a_i = 1$ for all frequencies $\nu_i$. Similarly, the vector $\vec{b}$ is the frequency dependence of the tSZ effect, and is therefore given by 
\be
b_i = f(\nu_i) \ , \ \ \ \  f(\nu_i) = x_i \coth{(x_i/2)} -4
\ee
and $x_i  = h\nu_i / k_B T_{CMB}$.  The ILC method solves for a set of weights $w_i$ such that the signal of interest is given by $\hat{s}(p) = w_i T_i(p)$, and that $\hat{s}(p)$ has minimum variance. Simple linear algebra shows that the solution is 
\be
\hat{s} = \frac{a_i R^{-1}_{ij}} {a_i R^{-1}_{ij} a_j} T_j \ \ \ \ {\rm and} \ \ \ \ \hat{y} = \frac{b_i R^{-1}_{ij}} {b_i R^{-1}_{ij} b_j} T_j
\label{eq:ILCweights}
\ee
where we have defined the covariance matrix of the observations $R_{ij} = \langle T_i T_j \rangle$. The matrix $R$ is usually evaluated empirically from the observations themselves. However, for the purpose of this paper, we will use a semi-analytical foreground model based on observations of the ACT experiment (\cite{2013JCAP...07..025D}) to evaluate $R$ at each multipole $\ell$ in harmonic space, as described below. 

The harmonic space foreground model is based on \cite{2013JCAP...07..025D} and includes contributions from the Cosmic Infrared Background (CIB, both clustered and Poisson components), extragalactic radio emission, galactic cirrus as well as tSZ and kSZ. We assume no decorrelation between frequencies, so that the template of each component is simply rescaled by frequency.  We also assume that the emission laws are known and independent of position in the sky, and we take fiducial values from Table 2 of \cite{2013JCAP...07..025D}. A slight generalization of Equation \ref{eq:ILCweights} allows us to include correlation between different components, such as CIB and tSZ. Given the uncertainty on this correlation at present, we chose to set it to zero in our fiducial model.

For each component and each multipole $\ell$ we evaluate the weights in Equation \ref{eq:ILCweights}. We compute the total power spectrum that a map would have when forming a linear combination with the given weights. We shall call the total power spectrum of the ILC map  $N^{{\rm ILC}, X}_\ell$ with $X = T$ for the cleaned CMB map and $X = y$ for the $y$ map. This notation is consistent with the fact that this total power spectrum will represent the ``noise" in our measurement of kSZ and tSZ from our tracer sample. Note that to be conservative we always include the `signal part' when estimating covariance matrices.

While in this work we have assumed no correlation between different components, this assumption is likely to fail at the level of precision required by the next generation experiments and constraints may be limited by our ability to model extragalactic foregrounds. Future mm and sub-mm experiments like CCAT-prime will provide higher frequency measurements compared to the ones considered in this work and are likely to improve our understanding and ability to remove foregrounds from lower frequency observations.

\begin{table*}
  \caption[CMB Experimental setup]{CMB Experimental setup}
  \label{tab:cmbexp}
  \begin{center}
   \leavevmode
   \begin{tabular}{c|c|c|c|c|c|c|c|c} 
     \hline \hline
     & \multicolumn{2}{c|}{Stage-3 CMB}   &  \multicolumn{2}{c|}{CMB-S4 (2.0$\arcmin$)}&  \multicolumn{2}{c|}{CMB-S4 (1.5$\arcmin$)}&  \multicolumn{2}{c}{CMB-S4 (1.0$\arcmin$)}\\
     \hline            
   Frequency & Beam & Noise RMS  & Beam & Noise RMS& Beam & Noise RMS& Beam & Noise RMS\\
      (GHz) &  (arcmin)& ($\mu$K-arcmin)  & (arcmin)& ($\mu$K-arcmin)  & (arcmin)& ($\mu$K-arcmin)  & (arcmin)& ($\mu$K-arcmin) \\
     \hline 
28 & 7.1 & 80 & 10.1 & 20.0 & 7.6 & 20.0 & 5.1 & 20.0 \\
41 & 4.8 & 70 & 6.9 & 17.5 & 5.1 & 17.5 & 3.4 & 17.5 \\
90 & 2.2 & 8 & 3.1 & 2.0 & 2.4 & 2.0 & 1.6 & 2.0 \\
150 & 1.3 & 7 & 2.0 & 1.8 & 1.5 & 1.8 &1.0 & 1.8\\
230 & 0.9 & 25 & 1.3 & 6.3 & 1.0 & 6.3 & 0.6 & 6.3 \\
     \hline
   \end{tabular}
  \end{center}
  \begin{quote}
    \noindent 
    For the Stage-3 CMB experiment sensitivities presented in this table we use the projections from \citet{AdvACT}; Table 2 in \citet{Srini2017} collects projections for other Stage-3 CMB experiments undertaking surveys for more context. As the surveys progress, these sensitivity estimates will mature. For example, the eventual map sensitivities will depend on the precise sky coverage and their uniformity \citep[e.g.,][]{FdB2016b}.
    \end{quote}
\end{table*}

\begin{figure*}
\begin{center}
  \hfill
  \resizebox{0.50\hsize}{!}{\includegraphics{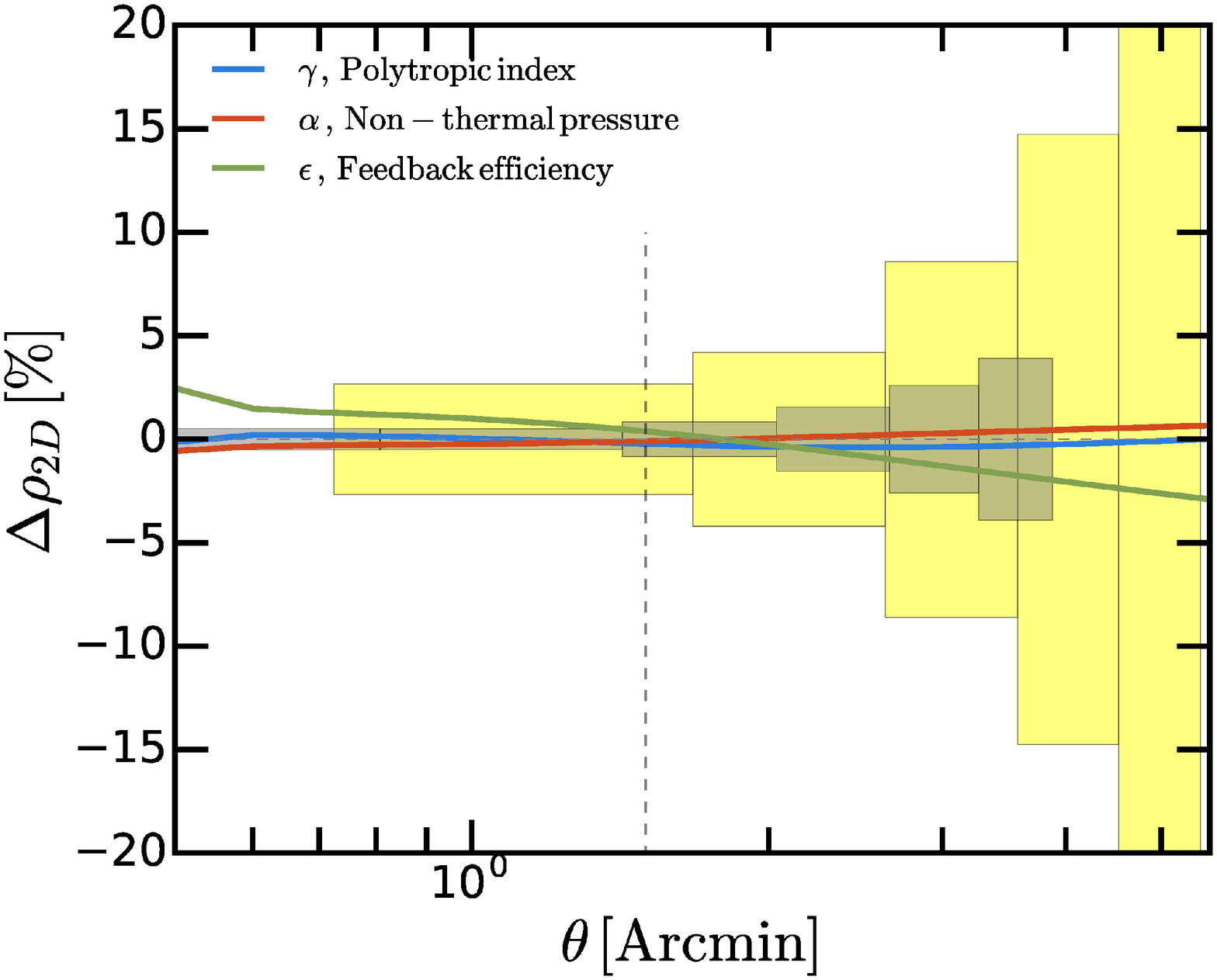}}%
  \resizebox{0.50\hsize}{!}{\includegraphics{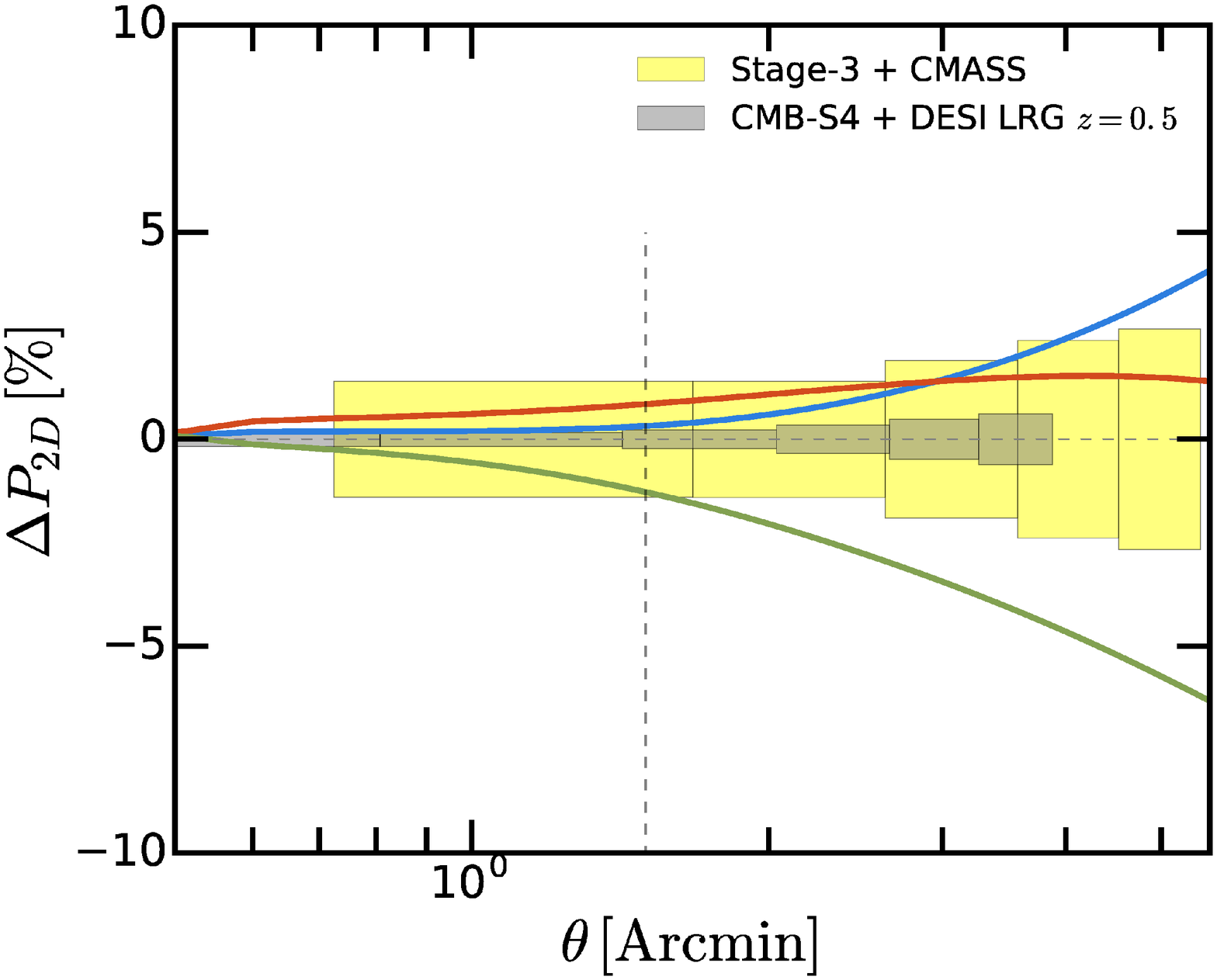}}\\
\end{center}
\caption{Shows the percent difference from the observed density (left panel) and pressure (right panel) profiles for a CMASS and LRG mass halo at $z=0.5$ when we vary the model parameters $\eps$, $\nth$, and $\Gamma$ around their fiducial values. The yellow and grey bands illustrate the forecasted errors for a Stage-3 CMB experiment cross correlated with CMASS and CMB-S4 cross correlated with DESI, respectively. These bands are a function of the beam size, so a smaller beam results in smaller observational bins closer to the halo center. The horizontal dashed lines illustrate the location of the virial radius for these halos. Stage-3 CMB experiments cross CMASS tSZ observations and CMB-S4 cross DESI kSZ observations (and tSZ) will be able to distinguish between the fiducial model and these variants.}
\label{fig:profs}
\end{figure*}

\subsection{Filtering and noise}
To extract information from a map (either CMB or $y$), we choose to use an Aperture Photometry (AP) filter of varying aperture $\theta_d$ (\cite{2015MNRAS.451.1606F}). Applying the filter consists in averaging the map values within a disk of radius $\theta_d$ and subtracting the mean temperature in an adjacent ring of equal area. More precisely, the output of the AP filter at the cluster location, hereafter labeled $A_X(\theta_d)$ (with $X = T, y$ for kSZ and tSZ respectively) is given by the following integral:
\be
A_X(\theta_d) = \int d^2 \theta \ \Psi_{\theta_d}(\theta) \ X(\vec{\theta})
\ee
where the AP filter $\Psi_{\theta_d}(\theta)$ is given by
\be
 \Psi_{\theta_d}(\theta) = \frac{1}{\pi \theta_d^2} \times \left\{
     \begin{array}{lr}
       \displaystyle
       1 & \, \mathrm{for}\  \, \theta < \theta_{d} \ , \\
       \displaystyle
      - 1 & \, \mathrm{for} \ \, \theta_d \leq \theta < \sqrt{2} \theta_d \ ,  \\
       \displaystyle
       0 & \mathrm{otherwise}.
     \end{array}
   \right.
\ee
A simple calculation shows that the error on $A_X$ due to primary CMB fluctuations as well as sources along the same line of sight (other than the one of interest) is given by
\bea
(M_X)_{\theta_d \theta_d'} & = & {\rm Cov}(A_X(\theta_d), A_X(\theta_d')) \nonumber \\ 
& = & \int \frac{d^2 \ell}{(2 \pi)^2} \Psi^*_{\theta_d}(\ell) \Psi_{\theta_d'}(\ell) N^{{\rm ILC}, X}_\ell
\eea
where $\Psi_{\theta_d}(\ell)$ is the 2D Fourier transform of $\Psi_{\theta_d}(\theta)$. Note that because of the shape of the filter, there is a non-trivial correlation between filters with different aperture, and this must be taken into account in the analysis.

\section{Results and Forecasts}
\label{sec:res}
The observations of the kSZ effect for a given sample currently are not significant enough to constrain thermodynamic properties for halos. The significance of kSZ measurements is expected to improve dramatically in the near future \citep[e.g.,][]{Flender2016,F16}. Looking ahead to these measurements, we forecast constraints on halo energetics for the parametric model presented in the previous section. We forecast the signal-to-noise and parameter constraints for Stage-3 and Stage-4 CMB experiments combined with the current and future spectroscopic surveys BOSS \citep{BOSS2011} and the Dark Energy Spectroscopic Instrument \citep[DESI,][]{DESI}, respectively. The proposed specifications for the CMB experiments are found in Table~\ref{tab:cmbexp}, including the next generation CMB-S4 experiment. We explore three different experimental designs for CMB-S4 that differ in their relative beam sizes, since the exact design of CMB-S4 is not finalized. The specifications for the spectroscopic surveys are found in Table~\ref{tab:desiexp}. We assume that the BOSS spectroscopic survey will have a 6000 square degree overlap with the Stage-3 CMB experiment and that a DESI-{\it like} experiment will have a 10000 square degree overlap with both CMB experiments.

We assign an average mass of $2\times10^{13}\,M_\odot$ when modeling the CMASS and LRG galaxy samples and an average mass of  $4\times10^{12}\,M_\odot$ when modeling the quasar (QSO) sample. When modeling the halos in the DESI survey, we choose to use the same average mass across all redshift bins and do not consider the distribution of masses in each bin. Since, we do not know what the mass distributions of the DESI survey samples will be, we have simplified the assumptions made about these samples and reduce this information down to average masses. When applying our method to observational data the mass distributions will need to be accounted for since the Compton-y signal is a non-linear function of mass.

The parameters that we vary in the ICM model are the efficiency of star formation feedback $\eps$, the amplitude of the non-thermal pressure profile $\nth$, and the polytropic index $\Gamma$. In Figure \ref{fig:profs} we show how each parameter affects the observed density and pressure profiles (left and right panels, respectively) for a CMASS and LRG mass halo at $z=0.5$. The solid lines show the difference in the observed profiles, $\Delta \rho_{2D} \equiv \rho_{2D,\rmn{new}}/ \rho_{2D,\rmn{fid}} - 1$ and $\Delta P_{2D} \equiv  P_{2D,\rmn{new}}/ P_{2D,\rmn{fid}} - 1$ when we change the model parameters by $\eps_\rmn{fid} - 1\times 10^{-6}$, $\nth _\rmn{fid} -  0.01$, and $\Gamma _\rmn{fid} - 0.01$ with respect to the fiducial model parameters $\eps = 2 \times 10^{-5}$, $\nth = 0.13$, and $\Gamma = 1.2$ for the given halo mass at $z=0.5$. Figure \ref{fig:profs} illustrates that decreasing feedback efficiency, $\eps$, increases the density profile in the inner regions of the halo, deceases it in the outer regions, and decreases the pressure profile on all scales, which is the results of the halo temperature profile being lower on all scales. Decreasing the amplitude of non-thermal, $\nth$, increases the pressure profile on all scales and does not impact the density profile at a significant level. The yellow and grey bands show the forecasted error bars on the observed density and pressure profiles on the fiducial model parameters at $z=0.5$ for a Stage-3 CMB experiment cross correlated with CMASS and CMB-S4 cross correlated with DESI, respectively.  It is clear that the differences between our fiducial parameters and their variants will be detectable with SZ observations with Stage-3 CMB experiments cross CMASS and SZ observations with CMB-S4 cross DESI. The differences shown here are the basis for numerical derivates used in the Fisher forecast. In these forecast we do not consider contributions from a 2-halo term \citep{Vinu2017}, although we are self consistent since we only model and fit for the 1-halo term. Additionally, we do not include the errors from the velocity reconstruction. The current state-of-the-art in velocity reconstruction on the CMASS sample has a correlation coefficient of around 0.7 \citep{Schaan2016}, which will decrease our forecasted kSZ signal-to-noise values proportional to the inverse of this coefficient. These errors from the velocity reconstruction methods are survey and technique dependent and they are expected to improve with higher density surveys and at higher redshifts. The velocity bias between halos with these masses and the reconstructed velocity field will be negligible and possibly detectable with greater than 100 $\sigma$ measurements, however, it will not affect the derivative of the density profile but only its amplitude.

\begin{figure*}`
 \begin{minipage}[t]{0.5\hsize}
    \centering{\small Stage-3 CMB + BOSS}
  \end{minipage}
  \begin{minipage}[t]{0.5\hsize}
      \centering{\small Stage-3 CMB + DESI}
  \end{minipage}
\begin{center}
  \hfill
  \resizebox{0.495\hsize}{!}{\includegraphics{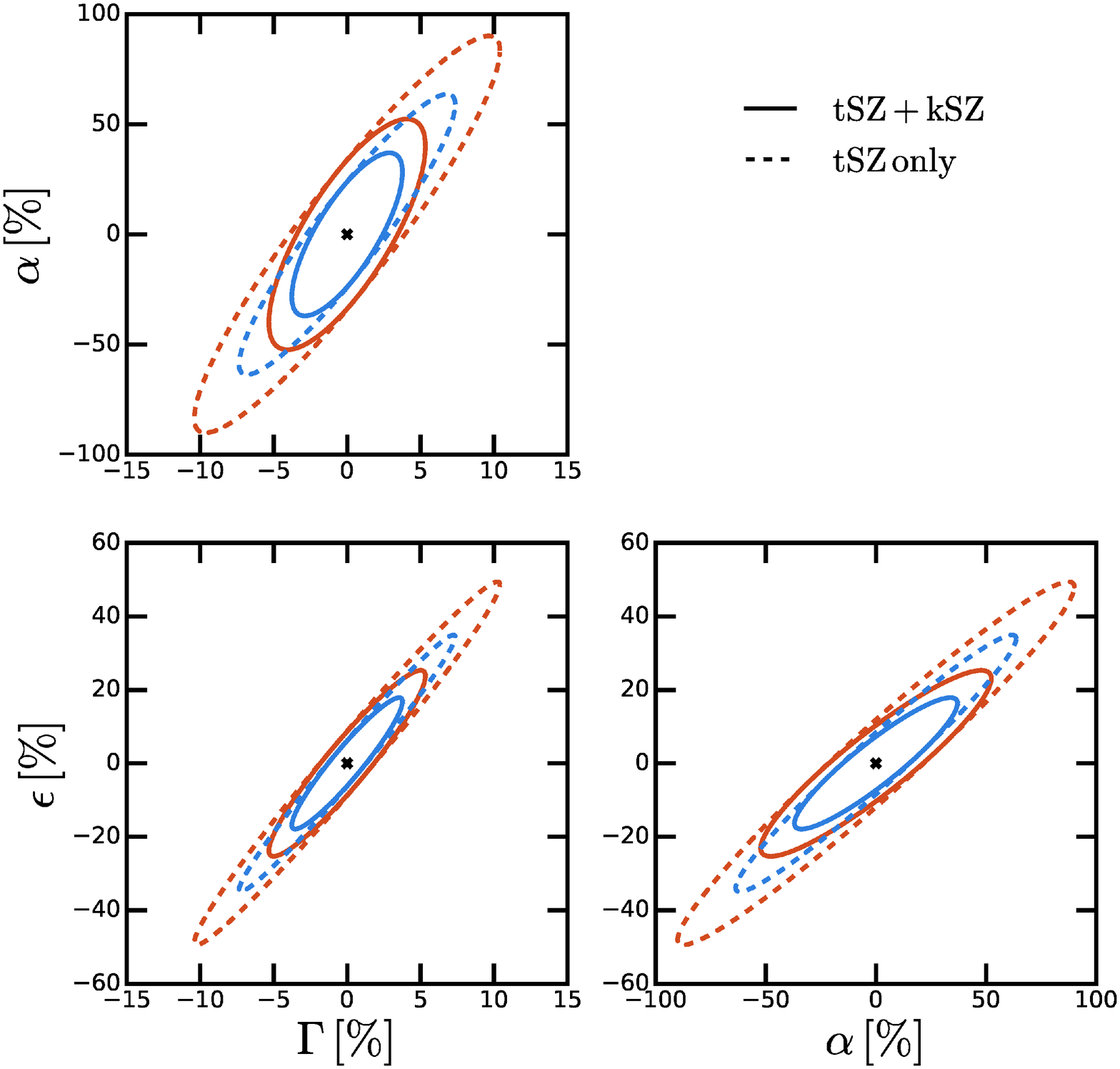}}                                                                                                          
  \resizebox{0.495\hsize}{!}{\includegraphics{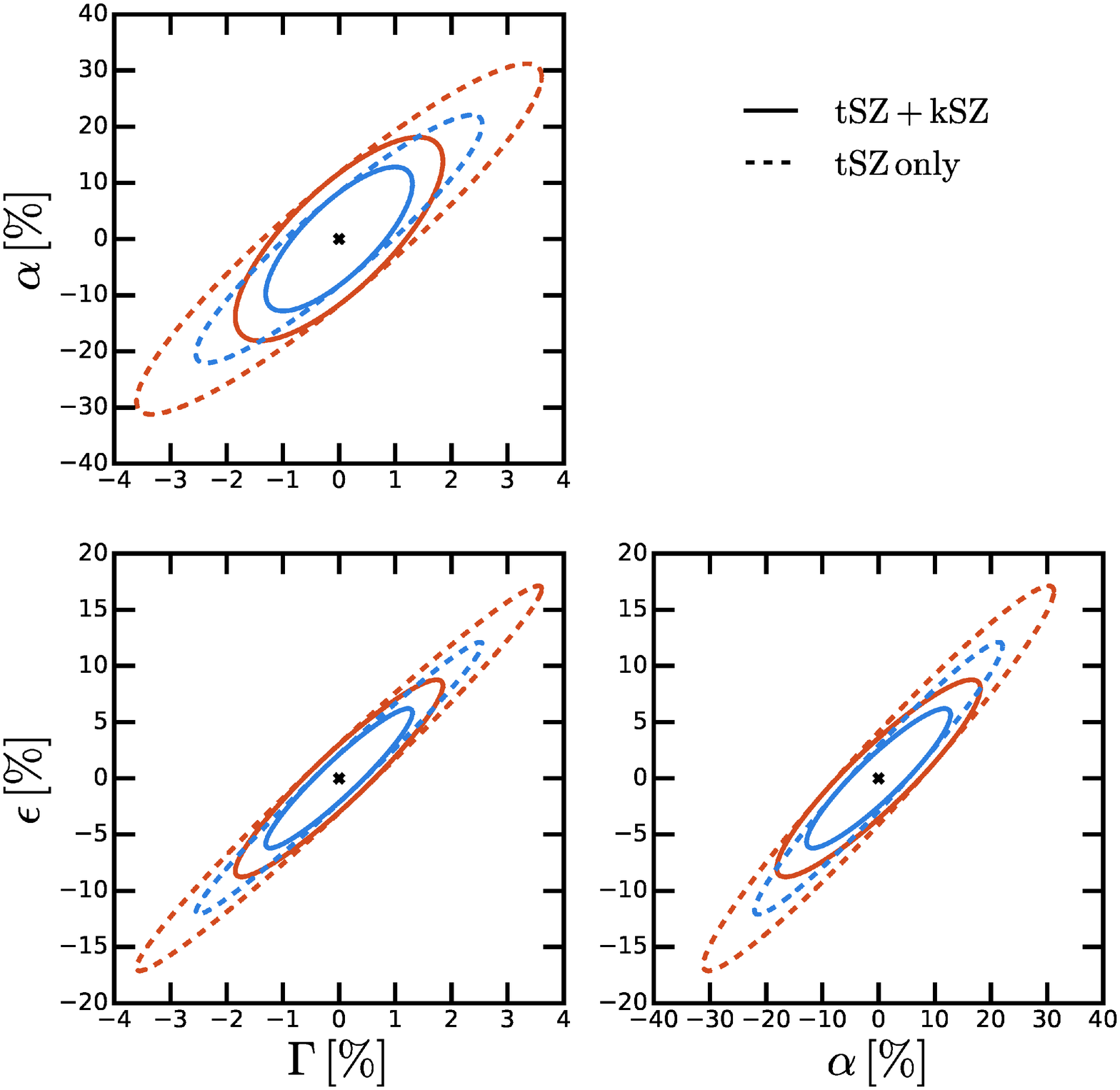}}\\
\end{center}
\caption{Fisher forecast constraints on $\eps$, $\nth$, and $\Gamma$ (percent errors) for the Stage-3 CMB experiment combined with a BOSS survey (left panel) and a DESI survey (right panel) survey for a CMASS and LRG samples, respectively. The solid lines show the constraints from combined tSZ and kSZ measurements and the dashed lines show the constraints only using tSZ measurements. The blue and red lines show the 68\% and 95\% error ellipses, respectively. The marginalized constraints on $\eps$ improves from 12\% to 4\% going from BOSS to DESI and similarly $\nth$ improves from 24\% to 8\%. Adding the kSZ measurements break degeneracies in all three parameters.}
\label{fig:fish}
\end{figure*}

\begin{table}
  \caption[DESI-like spectroscopic survey]{BOSS and DESI-like spectroscopic survey}
  \label{tab:desiexp}
  \begin{center}
   \leavevmode
   \begin{tabular}{c|c|c|c} 
     \hline \hline              
    & BOSS & \multicolumn{2}{c}{DESI} \\
    \hline
   $z$& CMASS [$10^4$]& $\dd N_\rmn{LRG}/\dd z^*\,[10^4]$ & $\dd N_\rmn{QS0}/\dd z^*\,[10^4]$  \\ 
     \hline 
0.1& - & 38 & 5\\
0.3& - & 126 & 22\\
0.5& 40 & 333 & 31 \\
0.7& - & 570 & 34 \\
0.9& - & 442 & 44 \\
1.1& - & 13 & 56 \\
1.3& - &  - & 69 \\
1.5& - & - &  81 \\
1.7& - & - & 80 \\
     \hline
   \end{tabular}
  \end{center}
  \begin{quote}
    \noindent 
$^*$ values for $\dd N/\dd z$ are from the DESI white paper \citep{DESI}.
The width of the redshift bins are 0.2 and we assume that for BOSS and DESI the overlapping area with CMB experiments is 6000 and 10000 square degrees, respectively. For the CMASS sample we assume a median redshift of 0.5.
  \end{quote}
\end{table}

We use the Fisher matrix formalism \citep[e.g.,][]{Fisher1935,Knox1995,Jung1996} to forecast the expected constraints on the parameters $\eps$, $\nth$, and $\Gamma$. The Fisher matrix $F_{jk}$ is calculated as

\be
F_{jk} =\sum_{X \in \{ T, y \}} \sum_{\theta_d, \theta_{d'}} \frac{\partial A_X(\theta_d)}{\partial p_j}  (M_{X}^{-1})_{\theta_d \theta_{d'}} \frac{\partial A_X(\theta_{d'})}{\partial p_k} 
\ee

\noindent where $A_X(\theta_d)$ is the output of the AP filter for the projected density or pressure profile aperture filter measurements, $(M_X^{-1})_{\theta_d \theta_{d'}}$ is the inverse covariance matrix for the aperture filter we are using and $p_j$ is $j^\rmn{th}$ parameter that we are forecasting. We use the full covariance matrix derived from the component separation described in Section \ref{sec:comp}, which includes estimates for the residual noise after foreground subtraction.

Figure~\ref{fig:fish} shows the fisher forecast constraints for halo samples from BOSS and DESI (left and right panel, respectively) on $\eps$, $\nth$, and $\Gamma$ cross correlated with a Stage-3 CMB experiment combined. The 1-$\sigma$ and 2-$\sigma$ error ellipses are shown by the blue and red lines, respectively. We illustrate how the kSZ observations of the density profile further constrain the IGM and ICM model parameters by showing the constraints with (solid lines) and without (dashed lines) the kSZ observations. The addition of kSZ observations helps break the degeneracy between all three parameters.  After marginalizing over the other parameters, we forecast that Stage-3 CMB experiments combined with BOSS will place 12\% constraints on $\eps$ for the CMASS sample and this will improve to 4\% with DESI for the LRG sample. Likewise we forecast that Stage-3 CMB experiments combined with BOSS will place 24\% constraints on $\nth$ for the CMASS sample and this will improve to 8\% with DESI for the LRG sample at $z=0.5$.

\begin{figure*}
 \begin{minipage}[t]{0.5\hsize}
    \centering{\small DESI LRG}
  \end{minipage}
  \begin{minipage}[t]{0.5\hsize}
    \centering{\small DESI QSO}
  \end{minipage}
\begin{center}
  \hfill
  \resizebox{0.50\hsize}{!}{\includegraphics{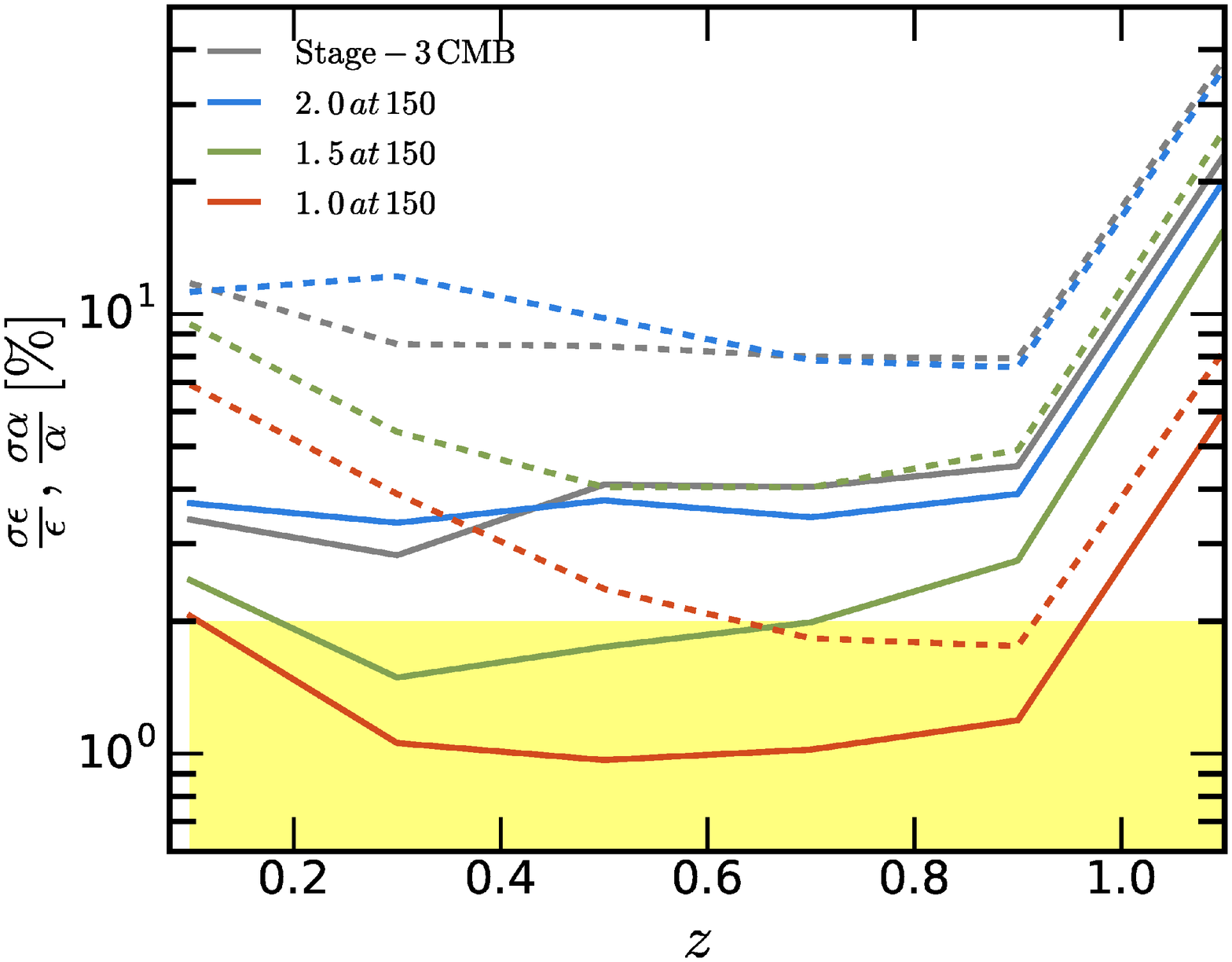}}%
  \resizebox{0.50\hsize}{!}{\includegraphics{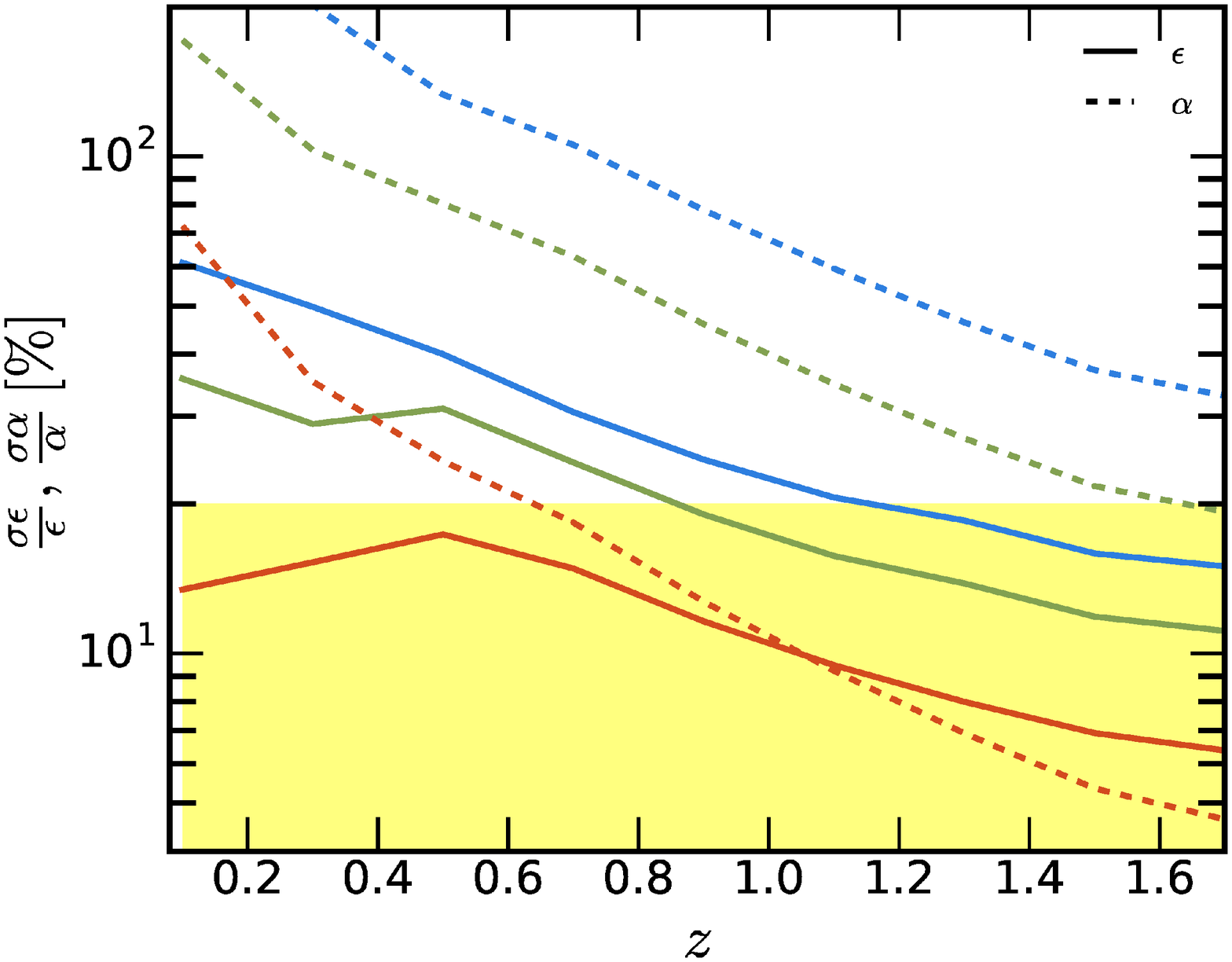}}\\
\end{center}
\begin{center}
  \hfill
  \resizebox{0.50\hsize}{!}{\includegraphics{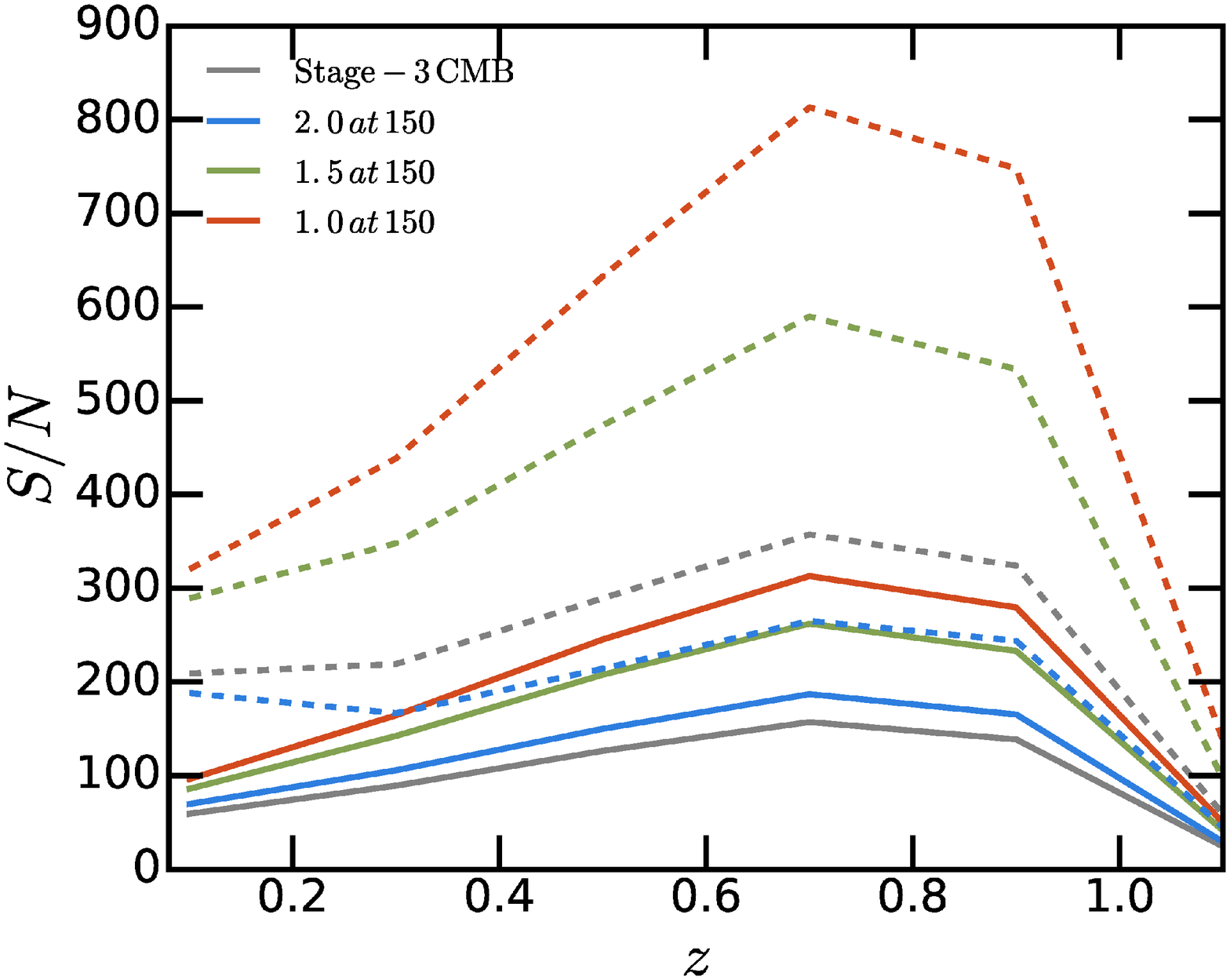}}%
  \resizebox{0.50\hsize}{!}{\includegraphics{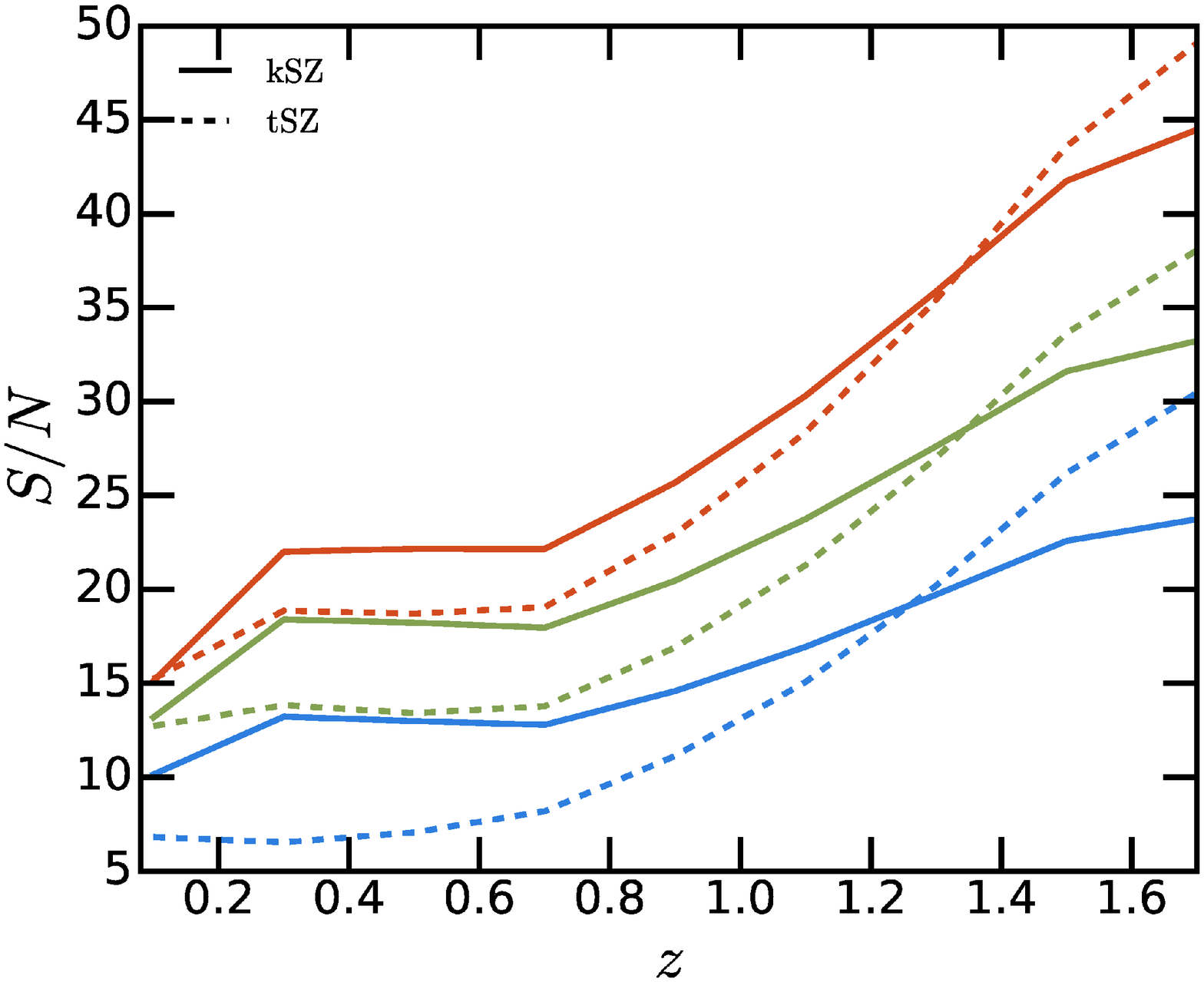}}\\
\end{center}
\caption{A resolution study on the beam size of a CMB-S4 at 150 GHz cross-correlated with LRG (left panels) and QSO (right panels) samples from a DESI-{\it like} survey using the marginalized constraints $\eps$ and $\nth$ (top panels) and the forecasted signal-to-noise ($S/N$) of tSZ and kSZ observations as a function of redshift. The marginalized constraints and $S/N$ are shown for a Stage-3 CMB experiment (grey lines), and CMB-S4 with a 1 (red lines), 1.5 (green lines), and 2 (blue lines) arcminute beams at 150 GHz. In the top panels the marginalized constraints are shown for $\eps$ with solid lines and for $\nth$ with dashed lines and the yellow shaded regions illustrate the 2\% and 20\% constraint thresholds. In the bottom panels the $S/N$ for the tSZ and kSZ observations are shown with solid and dashed lines, respectively. Clearly, a CMB-S4 with a 1 arcminute beam provides factors of 3 to orders of magnitude better constraints and $S/N$ compared to a 2 arcminute beam which is comparable to a Stage-3 CMB experiment. Regardless these forecasted constraints and $S/N$ will be an order of magnitude improvement over our current understanding of feedback efficiency and the thermodynamic properties of these halos at any redshift.}
\label{fig:fishcmb4}
\end{figure*}

Looking beyond Stage-3 CMB experiments, CMB-S4 will provide multi-band measurements in the microwave frequencies with increased sensitivity over a large fraction of the sky \citep{CMBS4}. Reaching these measurement goals will significantly increase the signal-to-noise of the cross correlations considered in this work and allow one to divide the galaxy or quasar sample as function of redshift (see Figure~\ref{fig:fishcmb4}). Beyond redshift, galaxy and quasar samples can be split by other classifications, such as halo mass, stellar mass, galaxy type etc. The main parameter that we consider here is the aperture which we define as the beam size at 150 GHz (see Table~\ref{tab:cmbexp}). Moreover, it is always possible to reduce the RMS noise level by integrating longer on the sky, whereas the aperture is fixed upon design completion.

In the top panels of Figure~\ref{fig:fishcmb4} we show the forecasted marginalized constraints on $\eps$ and $\nth$ for three possible CMB-S4 designs combined with DESI LRG (left panel) and DESI QSO (right panel) samples as a function of redshift. Additionally, for comparison, we show the constraints for a Stage-3 CMB experiment cross correlated with the DESI LRG sample. Clearly the forecasted constraints $\eps$ and $\nth$ depend strongly on the aperture of CMB-S4 with the constraints degrading by factors of 3 to larger than an order of magnitude for the DESI QSO sample. For the particular IGM and ICM model we used, 2\%  and 20\% constrains on thermal feedback efficiency are possible with CMB-S4 experiment that has a 1 arcminute beam at 150 GHz cross correlated with DESI LRG and QSO samples, respectively. Obtaining such constraints would be an order of magnitude improvement over our current understanding of feedback efficiency and the thermodynamic properties of these halos at any redshift. This is especially true for QSO samples where currently the tSZ signal is measured at $\sim 4 \sigma$ \citep{Ruan2015,Verdier2016,Crich2016,Soergel2016}. If CMB-S4 is built with a 2 arcminute beam at 150 GHz, we forecast that these constraints would show only marginal improvements or no improvements over Stage-3 CMB experiments together with DESI samples.

The results above are somewhat specific to our choice of model for the gas density and temperature. To remedy this, we also forecast the signal-to-noise ratio on the density and pressure profiles. These will be relevant when comparing any density and pressure profile from hydrodynamical simulation, and can therefore be turned into constraints on the simulation's sub-grid parameters. We show in the bottom panels of Figure~\ref{fig:fishcmb4} the raw signal-to-noise for CMB-S4 designs and a Stage-3 CMB experiment cross correlated with the DESI LRG (left panel) and DESI QSO (right panel) samples as a function of redshift. We forecast that the kSZ signal-to-noise from the QSO sample is larger than the tSZ signal-to-noise at $z < 1.3$. This results is expected given that the halo mass assigned to the QSO sample is $4\times10^{12}\,M_\odot$ and kSZ signal scales like halo mass compared the tSZ signal which scales like (halo mass)$^{5/3}$. The gains in signal-to-noise as a function of the beam size for CMB-S4 follow similar trends to the constraints on the IGM and ICM model parameters. Again, the 2 arcminute beam CMB-S4 experiment has comparable signal-to-noise to Stage-3 CMB experiments.

\section{Discussion and Conclusion}
\label{sec:con}
Motivated by the recent kSZ detections via cross-correlations and the promise of the large improvements in the signal-to-noise of future kSZ measurements, we propose the combination of kSZ and tSZ measurement to constraint global thermodynamic properties of the ionized baryons inside halos. In this work, we demonstrate that combined SZ measurements in the near future will constrain these properties, thus informing models on the complex processes involved in shaping the cosmic evolution of baryons and galaxies.

We forecasted the signal-to-noise and marginalized parameter constraints for Stage-3 CMB and CMB-S4 experiments combined with spectroscopic surveys such as BOSS and DESI, where the number densities of galaxies or QSO is sufficiently high such that the velocity reconstruction will not be shot-noise dominated. With Stage-3 CMB experiments and a DESI-{\it like} survey we forecast that one can constrain the average efficiency of feedback and amount of non-thermal pressure support in LRG halos at $z=0.5$ to better than 5\%. The high fidelity of these constraints combined with the improvements in sensitivity of future CMB experiments will allow us to further divide into sub-samples based on redshift or other classifications like galaxy or quasar type. Thus, with the signal-to-noise we forecast for CMB-S4 and DESI, one can study the evolution of energy injection and non-thermal pressure support in halos across cosmic time. The exact increase in signal-to-noise and parameters constraints going from Stage-3 CMB experiments to CMB-S4 will depend strongly on the beam size for CMB-S4. 

The SZ signal is redshift independent so we will extend this methodology to higher redshift objects like quasars, opening a new observational window into the thermodynamic properties of halos when the universe was a quarter of its age now. We forecast that for quasars selected in a DESI-{\it like} survey we can obtain $S/N > 25$ tSZ and kSZ observations from $z = 1.1 - 1.7$ with a CMB-S4 experiment that has a 1 arcminute beam. In principle similar analyses can be undertaken with other kSZ estimators like projected fields \citep{Hill2016,F16, 2016arXiv160701769S}. However, modeling the kSZ signal for such projected fields estimators is more involved and beyond the scope of this paper.

The constraints presented here provide complementary and observationally independent thermodynamic information on the ICM compared to X-ray observations of massive halos like clusters. We note that \citet{Flender2016b} constrains a very similar ICM model using X-rays observations, albeit with much larger errors. Currently, such analyses are done on pointed X-ray observations, however near future X-ray satellite  missions like eROSITA \citep{Erosita} and Athena \citep{Athena} will provided complementary observations of clusters and groups to the SZ observations used in this work, but currently only SZ observations can constrain the thermodynamic properties of the IGM for large samples.

Beyond the thermodynamic properties of halos, SZ observations will constrain the baryonic matter distribution and its effect on the matter power spectrum $P(k)$. For future cosmological probes, like cosmic shear, that push into the quasi-linear and non-linear regime of $P(k)$ these baryonic processes cannot be ignored and need to be modeled \citep[e.g.,][]{Semboloni2011,Eifler2015}. A complementary approach proposed in \citet{Foreman2016} uses cosmic shear measurements to indirectly probe baryonic processes through their effects on the matter power spectrum, but does not simultaneously constrain both cosmological parameters and baryonic processes. Exactly how the ICM parameter constraints and combination of SZ observations proposed in this work will map onto models for $P(k)$ are left to future work. The approaches proposed above will be complementary to previous models which calibrate baryonic effects into $P(k)$ using cosmological hydrodynamic simulations \citep[e.g.,][]{Mead2015,ST2015} and are essential for future cosmological experiments.

\acknowledgments

We thank Jim~Bartlett, Tom~Crawford, Colin~Hill, Eiichiro~Komatsu, Arthur~Kosowsky, Mathew~Madhavacheril, Mike~Niemack, Lyman~Page, Bjoern~Soergel, Suzanne~Staggs for their constructive comments and helpful discussions. NB acknowledges the support from the Lyman Spitzer Jr. Fellowship. SF is funded by the Miller Institute for Basic Research in Science at the University of California, Berkeley. ES acknowledges support from the National Science Foundation grant NSF AST-1311756. The Flatiron Institute is supported by the Simons Foundation.

\bibliography{nab}
\bibliographystyle{apj}

\appendix
\section{Virial theorem and gravitational potential energy}
\label{app:virial}
In this appendix, we give a general derivation of the virial theorem for gas with density $\rho_\text{gas}$ in a \textit{total} gravitational potential $\varphi$, valid even in the absence of spherical symmetry, and whether or not the potential is sourced by the gas alone.

We start by defining the quantity
\be
\Phi_\text{gas}
\equiv -
\int \dd^3\vec{r} \;
\rho_\text{gas} \;
\vec{r}\cdot \vec{\nabla}\varphi ,
\ee
Pressure equilibrium reads
$\vec{\nabla}P_\text{tot} = -\rho \vec{\nabla} \varphi$,
 and allows to rewrite:
 \be
 \bal
\Phi_\text{gas}
&=
\int \dd^3\vec{r} \;
\vec{r}\cdot \vec{\nabla} P_\text{tot}\\
&=
\int \dd^3\vec{r} \;
\left[
\text{div} (\vec{r} P_\text{tot})
-
P_\text{tot} \underbrace{\text{div} (\vec{r})}_{=3}
\right]\\
&=
- 3 \int \dd^3\vec{r} \; P_\text{tot}
+
\oiint d\vec{S}\cdot\vec{r} \; P_\text{tot},
\eal
\ee
or:
\be
\Phi_\text{gas}
+ 3 \int \dd^3\vec{r} \; P_\text{tot}
=
\underbrace{ \oiint d\vec{S}\cdot\vec{r} \; P_\text{tot} }
_\text{surface pressure}
,
\ee
which reduces to Equation~\eqref{eq:vir} in the spherically symmetric limit.

Let us now discuss the relation of $\Phi_\text{gas}$ to the gravitational energy of the gas. In the case of a self-gravitating gas, i.e. the gravitational potential $\varphi$ is sourced by the gas ($\nabla^2 \varphi = 4\pi G \rho_\text{gas}$), then $\Phi_\text{gas}$ is indeed the gravitational energy of the gas:
\be
\bal
\Phi_\text{gas}
&\equiv
-
\int \dd^3\vec{r} \;
\rho_\text{gas} \;
\vec{r}\cdot \vec{\nabla}\varphi\\
&= 
-\frac{1}{2}
 G
\int \dd^3\vec{r} \dd^3\vec{r}^\prime \;
\frac{\rho_\text{gas}(\vec{r}) \rho_\text{gas}(\vec{r}^\prime)}{ |\vec{r} - \vec{r}^\prime|}\\
&=
- G
\int \dd r \;
\rho_\text{gas}(r) \frac{M_\text{gas}(<r) }{r}
\hspace{0.5cm}\text{in spherical symmetry.}
\\
\eal
\ee

However, in our case where the potential $\varphi$ is not (entirely) sourced by the gas ($\nabla^2 \varphi = 4\pi G \rho_\text{tot}$), 
then $\Phi_\text{gas}$ is no longer the gravitational energy of the gas: 
\be
\bal
\Phi_\text{gas}
&\equiv
-
\int \dd^3\vec{r} \;
\rho_\text{gas} \;
\vec{r}\cdot \vec{\nabla}\varphi\\
&=
-G
\int \dd r \;
\rho_\text{gas}(r) \frac{M_\text{tot}(<r) }{r}
\hspace{0.5cm}\text{in spherical symmetry,}
\\
&\neq
\int \dd^3 \vec{r} \;
\rho_\text{gas}(\vec{r}) \varphi(\vec{r})
\hspace{0.5cm}\text{since}\hspace{0.5cm}
\varphi (r) = - G \frac{M_\text{tot}(<r)}{r}
- G \int_r^\infty \dd r^\prime \frac{M_\text{tot}(r^\prime)}{r^\prime}
\\
\eal
\ee
However, as explained in the main text, whether $\Phi_\text{gas}$ is rigorously the gravitational energy of the gas or not does not matter for our purpose, as long as it can be inferred from the gas profile (from kSZ) and the total mass distribution (NFW profile or measured from lensing).

\end{document}